\documentclass[a4paper,11pt]{article}
\usepackage{jcappub} % for details on the use of the package, please
\usepackage[T1]{fontenc}
\title{\boldmath Higher-order generalized uncertainty principle applied to gravitational baryogenesis}

\author[a,1]{Zhong-Wen Feng,\note{Corresponding author.}}
\author[a]{Xia Zhou,}
\author[b]{Shi-Qi Zhou}
\affiliation[a]{Physics and Space Science College, China West Normal University,\\Shida Road No.1, 637009 Nanchong China}
\affiliation[b]{School of Physics and Astronomy, Sun Yat-sen University, \\Daxue Road No. 2, 519082 Zhuhai China}
\emailAdd{zwfengphy@cwnu.edu.cn}

\abstract{The gravitational baryogenesis plays an important role in the study of baryon asymmetry. However, the original mechanism  of gravitational baryogenesis in the radiation-dominated era leads to the asymmetry factor $\eta$ equal to zero, which indicates this mechanism may not generate a sufficient baryon asymmetry in the early Universe. In this paper, we investigate the gravitational baryogenesis for the generation of baryon asymmetry in the early Universe by using a new higher-order generalized uncertainty principle (GUP). It is demonstrated that the entropy and the Friedman equation of the Universe deviate from the original cases due to the effect of the higher-order GUP. Those modifications break the thermal equilibrium of the Universe, and in turn produce a non-zero asymmetry factor $\eta $. In particular, our results satisfy all of Sakharov's conditions, which indicates that the scheme of explaining baryon asymmetry in the framework of higher-order GUP is feasible. In addition,
combining our theoretical results with the observational data, we constraint the GUP parameter $\beta_0$, whose bound is between $8.4 \times {10^{10}}  \sim 1.1 \times {10^{13}}$.}

\begin{document}

\maketitle
\flushbottom
\section{Introduction}
\label{intro}
The discovery of antimatter is one of the most important achievements in modern physics \cite{cha1}. Initially, it is thought that the number of matter and antimatter in the Universe was equal \cite{chb1}. However, this prediction conflicts with the observational evidence such as the Big Bang Nucleosynthesis (BBN), the Cosmic Microwave Background (CMB), and Planck observations, which show that matter exceeds antimatter \cite{cha2,cha3,cha3+}. This stark contradiction sparks an open issue of modern physics and cosmology, namely the baryon asymmetry of the Universe (BAU) \cite{cha4}.  Notably, there are two consensuses on the issue of baryon asymmetry, one is that baryon asymmetry may generate dynamically in the early stages of the cosmic expansion, and the other is that the key to the generation of baryon asymmetry is to satisfy the three Sakharov conditions, which are \cite{cha5} (i) existence of reactions violating baryon number; (ii) C (charge conjugation) and CP (combined charge conjugation and parity transformation) violation; (iii) departure from thermal equilibrium. Along this line, various theories have been proposed to explain how the asymmetry between baryons and antibaryons arose in the early stages of the evolution of the Universe (see \cite{cha6,cha11,chb11} for a review). In particular, Davoudiasl \emph{et al}. \cite{cha7} proposed one appealing mechanism called ``gravitational baryogenesis'', which is for generating the baryon number asymmetry in thermal equilibrium during the expansion of the Universe utilizing a dynamical breaking of CPT. The gravitational baryogenesis has gained people's attention as soon as it was put forward because this mechanism showed that gravitational coupling is a means to produce material asymmetry. Since then, various scenarios are in progress in this direction (see e.g.,~\cite{cha8,cha9,cha10,cha12,cha13}).

Although the gravitational baryogenesis plays an important role in analyzing BAU, it is found to be flawed in further research, i.e., in the general relativity regime for flat Friedmann-Robertson-Walker (FRW) metric, one found that the time changes of the Ricci scalar curvature  $\dot R$ and the corresponding baryon asymmetry factor  $\eta$ (a factor used to measure the number of baryonic matter exceeding the antibaryonic matter) are zero, which indicate the original mechanism cannot generate a sufficient baryon asymmetry for the radiation dominated universe. To address the problem above, many effective schemes have been proposed. For example, Li \emph{et al.} \cite{chb14} improved the interaction of gravitational baryogenesis to generate the matter-antimatter asymmetry. In~\cite{cha14,chx13,chx14,cha16,cha17,cha18, chxx19,chb18}, the authors argued that combining the original mechanism of gravitational baryogenesis with modified gravity theories can deviate the Friedmann equations from the classical case and produces a non-vanishing baryon asymmetry factor. Fukushima \emph{et al.} \cite{cha19} investigated the gravitational baryogenesis in the anisotropic spacetime and discussed the BAU at the end of anisotropic inflation. When considering the loop quantum cosmology effects, Odintsov and Oikonomou \cite{chx15} obtained a non-zero baryon-to-entropy ratio from the gravitational baryogenesis mechanism. Smyth \emph{et al}. \cite{chb19} discussed that the BAU is created through Hawking radiation from the primordial black holes via a dynamically-generated chemical potential.

On the other hand, a distinctive signature of most candidate theories of quantum gravity (QG) is the prediction of a minimum measurable length near the Planck scale \cite{cha20,cha21,cha22}. In that case, many classical theories should be modified by the effect of QG. For instance, by incorporating the minimal measurable length with the Heisenberg algebra, the conventional Heisenberg uncertainty principle (HUP) can be changed to the so-called generalized uncertainty principle (GUP). In recent years, the GUP has received wide attention since it can be used in physical systems with extremely small scale or high-energy scale \cite{cha23,cha24,cha25,cha26,cha27,cha28,cha29}. As discussed above, the BAU may occur at around the radiation domination era, during which the effect of GUP is also active at that time. In this connection, the combination of gravitational baryogenesis with GUP allows addressing the BAU. In~\cite{cha29+}, according to the GUP that was constructed by Ali, Das, and Vagenas (ADV model) \cite{cha30}, i.e.,  $\Delta x\Delta p \ge \frac{\hbar }{2}\left[ {1 - {{{\alpha _0}{\ell _p}\Delta p} \mathord{\left/ {\vphantom {{{\alpha _0}{\ell _p}\Delta p} \hbar }} \right. \kern-\nulldelimiterspace} \hbar }} \right.$ $\left. { + {{{\beta _0}\ell _p^2\Delta {p^2}} \mathord{\left/ {\vphantom {{{\beta _0}\ell _p^2\Delta {p^2}} {{\hbar ^2}}}} \right. \kern-\nulldelimiterspace} {{\hbar ^2}}}} \right]$ with the GUP parameters  ${\alpha _0}$ and ${\beta _0}$,\footnote{When ignoring the ${\alpha _0}$, the ADV model reduces to $\Delta x\Delta p \ge {{\hbar \left[ {1 + {{\beta _0\ell _p^2\Delta {p^2}} \mathord{\left/ {\vphantom {{\beta _0\ell _p^2\Delta {p^2}} {{\hbar ^2}}}} \right. \kern-\nulldelimiterspace} {{\hbar ^2}}}} \right]{\rm{ }}} \mathord{\left/ {\vphantom {{\hbar \left[ {1 + {{\beta _0\ell _p^2\Delta {p^2}} \mathord{\left/ {\vphantom {{\beta _0\ell _p^2\Delta {p^2}} {{\hbar ^2}}}} \right. \kern-\nulldelimiterspace} {{\hbar ^2}}}} \right]{\rm{ }}} 2}} \right. \kern-\nulldelimiterspace} 2}$, which is consistent with the  expression of KMM model.} Das \emph{et al}. derived a non-zero asymmetry factor $\eta$ at the radiation domination era, which provides a new way to explain BAU. However, it is worth mentioning that the ADV model has some limitations \cite{cha31}. Firstly, the perturbative of the ADV model is only valid for small values of the GUP parameter. Secondly, this GUP model does not imply noncommutative geometry. In addition, due to the maximal momentum uncertainty being different from the maximal momentum, it is found that the ADV model is not appropriate to the doubly special relativity. To solve those defects, Pedram presented a nonperturbative higher-order GUP that is in agreement with various proposals of QG and eliminates the objections due to the doubly special relativity theories \cite{cha32}.

This heuristic work of Pedram is considered as an important step towards establishing a physically reasonable theory of QG.  Therefore, many studies have converged on the construction and application of high-order GUP \cite{cha33,cha34,cha35,chc36}. Recently, according to the commutator  $\left[ {x,p} \right] = i\hbar \sqrt {1 - 2\beta {p^2}} $, Petruzziello \cite{cha36} constructed a new higher-order GUP, viz.
\begin{align}
\label{eq1}
\Delta x\Delta p & \ge \frac{\hbar }{2}\left\langle {\sqrt {1 - 2\beta {p^2}} } \right\rangle \ge \frac{\hbar }{2}\sum\limits_{n = 0}^\infty  {\frac{{{\beta ^n}\left( {2n} \right)!}}{{{2^n}\left( {1 - 2n} \right){{\left( {n!} \right)}^2}}}} \left\langle {{p^{2n}}} \right\rangle
\nonumber \\
& \ge \frac{\hbar }{2}\sum\limits_{n = 0}^\infty  {\frac{{{\beta ^n}\left( {2n} \right)!}}{{{2^n}\left( {1 - 2n} \right){{\left( {n!} \right)}^2}}}} {\left\langle {{p^2}} \right\rangle ^n}
\nonumber \\
& \ge \frac{\hbar }{2}\sum\limits_{n = 0}^\infty  {\frac{{{\beta ^n}\left( {2n} \right)!}}{{{2^n}\left( {1 - 2n} \right){{\left( {n!} \right)}^2}}}} {\left( {\Delta {p^2} + {{\langle \hat p\rangle }^2}} \right)^n}
 \nonumber \\
& \ge \frac{\hbar }{2}\sqrt {1 - 2\beta \Delta {p^2}},
\end{align}
where we have used the property  $\left\langle {{p^{2n}}} \right\rangle  \ge {\left\langle {{p^2}} \right\rangle ^n}$ and the mirror-symmetric states  $\left\langle p \right\rangle  = 0$, $\beta  = {{{\beta _0}\ell _p^2} \mathord{\left/ {\vphantom {{{\beta _0}\ell _p^2} {{\hbar ^2}}}} \right. \kern-\nulldelimiterspace} {{\hbar ^2}}} = {{{\beta _0}} \mathord{\left/ {\vphantom {{{\beta _0}} {m_p^2{c^2}}}} \right. \kern-\nulldelimiterspace} {m_p^2{c^2}}}$ with the dimensionless GUP parameter  $\beta_0$, the Planck length  ${\ell _p}$ and the Planck mass ${m _p}$. In addition to inheriting the advantages of the previous higher-order GUP, this model also has the following characteristics: (i) it has a negative value for the GUP parameter  $- \left| \beta_0  \right|$, which is considered to be very useful in astrophysical and cosmology. In~\cite{chb36}, it is estimated that the negative GUP parameter is the only setting meets the Chandrasekhar limit for white dwarfs. Besides, it is found that the negative GUP parameter may appears in non-trivial space-time structures and leads to a crystal-like universe \cite{chc37,chb37}.
For the sake of simplicity, we omit absolute value of the GUP parameter in this work; (ii) eq.~(\ref{eq1}) contains only a maximal observable momentum  $\Delta {p_{\max }} \approx {1 \mathord{\left/ {\vphantom {1 {\sqrt {2\beta } }}} \right. \kern-\nulldelimiterspace} {\sqrt {2\beta } }}$ and no minimal length uncertainty, which is never occurs in other models. For $\Delta p =  \pm \Delta {p_{\max }}$, the position and momentum operators of above inequality are interchangeable, which is consistent with the classical case; (iii) this new higher order GUP reduce to the quadratic form of the  GUP $\Delta x\Delta p \ge {{\hbar \left[ {1 + {{\beta _0^{{\rm{KMM}}}\ell _p^2\Delta {p^2}} \mathord{\left/ {\vphantom {{\beta _0^{{\rm{KMM}}}\ell _p^2\Delta {p^2}} {{\hbar ^2}}}} \right. \kern-\nulldelimiterspace} {{\hbar ^2}}}} \right]{\rm{ }}} \mathord{\left/ {\vphantom {{\hbar \left[ {1 + {{\beta _0^{{\rm{KMM}}}\ell _p^2\Delta {p^2}} \mathord{\left/ {\vphantom {{\beta _0^{{\rm{KMM}}}\ell _p^2\Delta {p^2}} {{\hbar ^2}}}} \right. \kern-\nulldelimiterspace} {{\hbar ^2}}}} \right]{\rm{ }}} 2}} \right. \kern-\nulldelimiterspace} 2}$   (hereafter referred to as Kempf-Mangano-Mann model or ``KMM model'') with the negative GUP parameter ${\beta_0 ^{{\rm{KMM}}}}$ for  $\sqrt \beta  \Delta p \ll 1$. To sum up, it is found that the GUP parameters in the two models have has the  relationship  ${\beta_0}  \sim  - {\beta_0 ^{{\rm{KMM}}}}$. However, when the GUP parameters equal to zero, they recover the HUP.

In \cite{chy1,chy2}, the authors pointed out that features of QG are imprinted on evolution of the Universe, which can be decoded through the astronomical observation (such as CMB \cite{chy3,chy4} and the upcoming Cherenkov Telescope Array (CTA) \cite{cha52}). Furthermore, the effect of QG in the higher-order proposal would greatly change the classical physical system. Therefore, studying the BAU in the framework of higher-order GUP helps people understand the role of QG in the early evolution of the Universe. However, to the best of our knowledge, the higher-order GUP applied to baryogenesis has never been reported. To address this issue, in this present paper we attempt to employ the new higher-order GUP to study the BAU. We intend to combine eq.~(\ref{eq1}) with the gravitational baryogenesis to derive the modified Friedmann equations, which in turn produces a non-zero time derivative of the Ricci scaler curvature $\dot R$ as well as a non-zero asymmetry factor  $\eta $. Then, the BAU can be explained due to those modifications. Finally, according to the recent observation data, it is possible to estimate the bound of the GUP parameter ${\beta _0}$. According to the unique properties of the new higher-order GUP, it is believed our study would give some different results from the previous works.

 This paper is organized as follows. In the next section, we give a brief review of the gravitational baryogenesis. In Section~\ref{sec3}, we derive the modified entropy and the corresponding Friedmann equations in the framework of the new higher-order GUP~(\ref{eq1}). Then, we derive the non-zero Ricci scalar curvature $R$  and the asymmetry factor $\eta$. Finally, the BAU is explained due to those modifications. In Section~\ref{sec4}, according to the recent experiments and observations, we constrain the bounds of the GUP parameter  $\beta_0$. The paper ends with conclusions in Section~\ref{sec5}. To simplify the notation, this research takes the  units $\hbar=c={k_B}=1$.

\section{Gravitational baryogenesis for the standard cosmological model}
\label{sec2}
Within supergravity theories, the interaction of the gravitational baryogenesis between the derivative of the Ricci scalar curvature and the baryon-number current dynamically breaks CPT in an expanding universe is given by~\cite{cha7}
\begin{align}
\label{eq2}
\frac{1}{{M_*^2}}\int {{{\rm{d}}^4}} x\sqrt { - g} \left( {{\partial _\mu }R} \right){J^\mu },
\end{align}
where  $R$ denotes the Ricci scalar curvature,  ${J^\mu }$ is the baryon current, ${M_*}$  is the cut-off scale of the effective theory. Following the viewpoint in~\cite{cha14,chx13,chx14,cha16,cha17,cha18, chxx19,chb18,cha19,cha37}, in the process of the Universe's expansion, when the temperature $T$  of the Universe drops below the critical temperature  $T_D$ at which the baryon asymmetry generating interactions occurs, the baryon asymmetry factor (BAF)  $\eta$  that measure the amount of baryon matter exceeds antibaryon matter can be derived from the equation above
\begin{align}
\label{eq3}
\eta  = \frac{{{n_B}}}{s} \simeq  - {\left. {\frac{{15{g_b}}}{{4{\pi ^2}{g_*}}}\frac{{\dot R}}{{M_*^2T}}} \right|_{{T_D}}},
\end{align}
where  ${{{n_B}} \mathord{\left/ {\vphantom {{{n_B}} s}} \right. \kern-\nulldelimiterspace} s}$ is the baryon to entropy ratio with the number of baryons (antibaryons) per volume unity  ${n_B}$ and the entropy density $s = {{2{\pi ^2}{g_*}{T^3}} \mathord{\left/ {\vphantom {{2{\pi ^2}{g_*}{T^3}} {45}}} \right. \kern-\nulldelimiterspace} {45}}$ for the Universe.  $\dot R = {{\partial R} \mathord{\left/ {\vphantom {{\partial R} {\partial t}}} \right. \kern-\nulldelimiterspace} {\partial t}}$ is the time derivative of the Ricci scalar curvature of the Universe,  $g_b$ and ${g_*}$  stand for the number of intrinsic degrees of freedom of baryons and the degrees of freedom of particles that contribute to the entropy of universe, respectively. According to eq.~(\ref{eq3}), one can observe that the BAF is proportional to $\dot R$. In the standard cosmological mode, the Ricci scalar curvature can be expressed as
\begin{align}
\label{eq4}
R =  - 8\pi G\left( {\rho  - 3p} \right),
\end{align}
with the energy density  $\rho$ and pressure  $p$ of the Universe. When considering matter source in the universe as a perfect fluid, one has the equation of state parameter $w = {p \mathord{\left/ {\vphantom {p \rho }} \right. \kern-\nulldelimiterspace} \rho }$, and the Eq.~(\ref{eq4}) can be cast as
\begin{align}
\label{eq5}
R =  - 8\pi G(1 - 3w)\rho.
\end{align}
Notably, when taking into account the radiation-dominated era that is characterized by $w = {1 \mathord{\left/ {\vphantom {1 3}} \right. \kern-\nulldelimiterspace} 3}$, the Ricci scalar curvature $R$ and its derivative  $\dot R$  vanish. As a consequence, one has $\eta=0$. However, many astronomical observations demonstrated the BAF is not equal to zero, which implies the matter in the Universe is more than the antimatter. For example, the acoustic peaks in CMB measured by the Wilkinson Microwave Anisotropy Probe (WMAP) give  $\eta  \le 6.3 \times {10^{ - 10}}$, BBN exhibits that $3.4 \times {10^{ - 10}} \le \eta  \le 6.9 \times {10^{ - 10}}$. To fix this contradiction, we investigate the higher-order GUP corrected gravitational baryogenesis in the next section.

\section{Gravitational baryogenesis in the framework of higher-order GUP}
\label{sec3}
In~\cite{cha29+}, the authors pointed out that, the preservation of thermal equilibrium in the original scenario of gravitational baryogenesis will lead to the Ricci scalar curvature as well as its derivative equal to zero, and eventually make  $\eta=0$. Therefore, an effective way to generate the baryon asymmetry is to break the thermal equilibrium of the Universe by deviating its thermodynamic properties from the classical case. In the following discussion, according to the holographic principle and the new higher-order GUP~(\ref{eq1}), we will derive the modified Bekenstein-Hawking entropy and the modified Friedmann equations, which break the thermal equilibrium. Finally, according to those modifications,  a non-zero Ricci scalar curvature  $R$ and a non-zero  $\eta $ can be obtained.

\subsection{The modified entropy in the higher-order GUP}
\label{sec3.1}
According to the holographic principle, when a gravitational system  absorbs a particle, its area and the total energy inside the apparent horizon will increases, the minimal change of the area $\Delta A$ can be expressed as \cite{cha38,chb34}
\begin{align}
\label{eq6}
\Delta A \sim Xm,
\end{align}
where  $X$ and $m$  represent the size and mass of the particle, respectively. In quantum mechanics, the width of the wave packet of a particle is described as the standard deviation of  $X$ distribution (i.e., the position uncertainty  $\Delta x$), hence, one has the relationship  $X \sim \Delta x$. Furthermore, in the process of measuring the position of a particle,  the mass of particle should be larger the momentum uncertainty  $\Delta p$ \cite{cha35}. Hence, eq.~(\ref{eq6}) can be rewritten as
\begin{align}
\label{eq7}
\Delta A \geq \Delta x\Delta p.
\end{align}
The above equation implies that the smallest increase in area in a gravitational system is restricted by the momentum uncertainty  $\Delta p$ and position uncertainty  $\Delta x$  of quantum mechanics.

For a static and spherically gravitational system, the position uncertainty is approximately equal to the radius of the apparent horizon, that is  $\Delta x \approx 2r$ \cite{chb38}. Substituting this relationship into the higher-order GUP~(\ref{eq1}), one has
\begin{align}
\label{eq8}
\frac{{\sqrt {1 - 2\Delta {p^2}\beta } }}{{2\Delta p}} \le \Delta x \le 2r,
\end{align}
and the momentum uncertainty is given by
\begin{align}
\label{eq9}
 - \frac{1}{{\sqrt {4\Delta {x^2} + 2\beta } }} \le \Delta p \le \frac{1}{{\sqrt {4\Delta {x^2} + 2\beta } }}.
\end{align}
Inserting eq.~(\ref{eq8}) and eq.~(\ref{eq9}) into eq.~(\ref{eq7}), the minimal change of the area can be rewritten as follows
\begin{align}
\label{eq10}
\Delta A \ge \chi \tilde \hbar \left( \beta  \right),
\end{align}
with the effective Planck constant ${\tilde \hbar} \left( \beta  \right) = {{2r} \mathord{\left/ {\vphantom {{2r} {\sqrt {16{r^2} + 2\beta } }}} \right. \kern-\nulldelimiterspace} {\sqrt {16{r^2} + 2\beta } }}$ and the calibration factor $\chi  = 4\ln 2$ \cite{chb39}. Notedly, in the limit $\beta\rightarrow0$, one yields ${\tilde \hbar} \left( \beta  \right)=1/2$. Based on the information theory, the minimal increase of entropy is conjectured related to the value of the area
\begin{align}
\label{eq11}
\frac{{{\rm{d}}S}}{{{\rm{d}}A}}\simeq\frac{{\Delta {S_{\min }}}}{{\Delta {A_{\min }}}} = \frac{1}{{8 {\tilde \hbar}  \left( \beta  \right)}},
\end{align}
where ${\Delta {S_{\min }}}= \ln 2$ represents the minimal increase in entropy, which implies to one bit of information. In the classical limit, the original entropy of a gravitational system can be expressed as ${S_0} = {A \mathord{\left/ {\vphantom {A 4}} \right. \kern-\nulldelimiterspace} 4}$, which indicates the entropy is related to the horizon area $A$. However, when considering the effect of GUP, the general expression of entropy should be modified as $S = {{f\left( A \right)} \mathord{\left/ {\vphantom {{f\left( A \right)} 4}} \right.  \kern-\nulldelimiterspace} 4}$ with  the  function of area  ${f\left( A \right)}$. Correspondingly, the entropy-area relation can be expressed as follows \cite{cha35+}
\begin{align}
\label{eq12}
\frac{{{\rm{d}}S}}{{{\rm{d}}A}}{\rm{ = }}\frac{{f'\left( A \right)}}{4},
\end{align}
where $f'\left( A \right) = {{{\rm{d}}f\left( A \right)} \mathord{\left/ {\vphantom {{{\rm{d}}f\left( A \right)} {{\rm{d}}A}}} \right. \kern-\nulldelimiterspace} {{\rm{d}}A}}$. By comparing eq.~(\ref{eq12}) with eq.~(\ref{eq11}), one has
\begin{align}
\label{eq13}
f'\left( A \right) = \frac{1}{{2 {\tilde \hbar} \left( \beta  \right)}} = \sqrt {1 + \frac{{\pi \beta }}{{2A}}}.
\end{align}
For  $\beta  \to 0$, one has  $f'\left( A \right) = 1$, consistently with  the standard result in classical limit. Then, by integrating eq.~(\ref{eq12}), the GUP corrected entropy is given by
\begin{align}
\label{eq14}
{S_{{\rm{GUP}}}} & = \int {\frac{{{\rm{d}}S}}{{{\rm{d}}A}}{\rm{d}}A}  \simeq \int {\frac{{\Delta {S_{\min }}}}{{\Delta {A_{\min }}}}{\rm{d}}A}  \simeq \int {\frac{{{\rm{d}}A}}{{8 {\tilde \hbar} \left( \beta  \right)}}}
\nonumber \\
& =\frac{{\sqrt {2\left( {2A + \pi \beta } \right)A} }}{8} + \frac{\pi }{8}\beta \ln \left( {2\sqrt A  + \sqrt {4A + 2\pi \beta } } \right).
\end{align}
Obviously, the effect of  GUP leads to a logarithmic term with the deformation parameter  $\beta$ in the parentheses of eq.~(\ref{eq14}), which meet the requirements of QG \cite{cha38,cha39,cha40}. Besides, in the limit where  $\beta  \to 0$, it agrees with the original entropy  ${S_0}$.
\subsection{GUP corrected Friedmann equations}
\label{sec3.2}
In~\cite{cha41,cha42,cha43,cha44}, it is found that the Friedmann equations can be derived from the Bekenstein-Hawking entropy and the first law of thermodynamics. Therefore, according to the modified entropy~(\ref{eq14}), we will derive the GUP corrected Friedmann equations that deviates from the thermal equilibrium. In the homogeneous and isotropic spacetime, the FRW universe is described by the line element:
\begin{align}
\label{eq15}
{\rm{d}}{s^2} = {h_{\mu \nu }}{\rm{d}}{x^\mu }{\rm{d}}{x^\nu } + {\tilde r^2}\left( {\rm{d}{\theta ^2} + {{\sin }^2}\theta \rm{d}{\varphi ^2}} \right),
\end{align}
where  ${x^\mu } = \left( {t,r} \right)$,  $\tilde r = ra\left( t \right)$ with the scale factor  $a\left( t \right)$, ${h_{\mu \nu }} = {\mathop{\rm diag}\nolimits} \left[ { - 1,{{{a^2}} \mathord{\left/ {\vphantom {{{a^2}} {\left( {1 - k{r^2}} \right)}}} \right. \kern-\nulldelimiterspace} {\left( {1 - k{r^2}} \right)}}} \right]$  is the two-dimensional metric with the spatial curvature constant  $k$, and  $\mu  = \nu  = 0,1$, respectively. By using the relation ${h^{\mu \nu }}{\partial _u}\tilde r{\partial _\nu }\tilde r = 0$, the dynamical apparent horizon of the FRW universe reads  $\tilde r = ar = {\left( {{H^2} + {k \mathord{\left/ {\vphantom {k {{a^2}}}} \right. \kern-\nulldelimiterspace} {{a^2}}}} \right)^{ - \frac{1}{2}}}$ with the Hubble parameter $H = {{\dot a} \mathord{\left/ {\vphantom {{\dot a} a}} \right. \kern-\nulldelimiterspace} a}$. By supposing the matter of the FRW universe is a perfect fluid, the energy-momentum tensor can be expressed as
\begin{align}
\label{eq16}
{T_{\mu \nu }} = \left( {\rho  + p} \right){u_\mu }{u_\nu } + p{g_{\mu \nu }},
\end{align}
where ${u_\mu }$ and ${g_{\mu \nu }}$  are the four velocity of the fluid and the space-time metric of FRW universe, respectively. $\rho_0$ is the energy density, and $p$ denotes the pressure. Due to the conservation law of energy-momentum  $T_{;\nu }^{\mu \nu } = 0$, one obtains the continuity equation, i.e.,
\begin{align}
\label{eq17}
\dot \rho  + 3H(\rho  + p) = 0.
\end{align}
Now, following the viewpoint in~\cite{cha45}, we will use the first law of thermodynamics for the matter content within the apparent horizon to derive the Friedmann equations. The first law of thermodynamics can be expressed as
\begin{align}
\label{eq18}
{\rm{d}}E = T{\rm{d}}S + W{\rm{d}}V,
\end{align}
where $E = \rho V$ represents the total energy of matter contained in the apparent horizon, $V = \left( {{4 \mathord{\left/ {\vphantom {4 3}} \right. \kern-\nulldelimiterspace} 3}} \right)\pi {\tilde r^3}$  is the volume of 3-dimensional sphere, and $W = {{\left( {\rho  - p} \right)} \mathord{\left/ {\vphantom {{\left( {\rho  - p} \right)} 2}} \right. \kern-\nulldelimiterspace} 2}$  is the work density, respectively. Based on the continuity equation~(\ref{eq17}), the energy differential turns out to be
\begin{align}
\label{eq19}
{\rm{d}}E = \rho {\rm{d}}V + V{\rm{d}}\rho  = 4\pi {\tilde r^2}\rho {\rm{d}}\tilde r + \frac{{4\pi {{\tilde r}^3}}}{3}{\rm{d}}\rho .
\end{align}
Then, considering the temperature of the apparent horizon $T = {\kappa  \mathord{\left/ {\vphantom {\kappa  {2\pi }}} \right. \kern-\nulldelimiterspace} {2\pi }}$  with the surface gravity for the metric FRW universe  $\kappa  = {{ - \left[ {1 - {{\left( {{{\partial \tilde r} \mathord{\left/ {\vphantom {{\partial \tilde r} {\partial t}}} \right. \kern-\nulldelimiterspace} {\partial t}}} \right)} \mathord{\left/ {\vphantom {{\left( {{{\partial \tilde r} \mathord{\left/ {\vphantom {{\partial \tilde r} {\partial t}}} \right. \kern-\nulldelimiterspace} {\partial t}}} \right)} {2H\tilde r}}} \right. \kern-\nulldelimiterspace} {2H\tilde r}}} \right]} \mathord{\left/ {\vphantom {{ - \left[ {1 - {{\left( {{{\partial \tilde r} \mathord{\left/ {\vphantom {{\partial \tilde r} {\partial t}}} \right. \kern-\nulldelimiterspace} {\partial t}}} \right)} \mathord{\left/ {\vphantom {{\left( {{{\partial \tilde r} \mathord{\left/ {\vphantom {{\partial \tilde r} {\partial t}}} \right. \kern-\nulldelimiterspace} {\partial t}}} \right)} {2H\tilde r}}} \right. \kern-\nulldelimiterspace} {2H\tilde r}}} \right]} {\tilde r}}} \right. \kern-\nulldelimiterspace} {\tilde r}}$ and the expression of entropy~(\ref{eq14}), the first term on the right hand side of eq.~(\ref{eq18}) can be rewritten as
\begin{align}
\label{eq20}
T{\rm{d}}S =  - \frac{1}{G}\left( {1 - \frac {{\dot{\tilde r}}}{{2H\tilde r}}} \right)f'\left( A \right),
\end{align}
where we denoted  $\dot{\tilde r}= {{\partial \tilde r} \mathord{\left/ {\vphantom {{\partial \tilde r} {\partial t}}} \right. \kern-\nulldelimiterspace} {\partial t}}$. Furthermore, the second term on the right hand side of eq.~(\ref{eq18}) is evaluated by
\begin{align}
\label{eq20+}
W{\rm{d}}V = 2\pi {\tilde r^2}\left( {\rho  - p} \right){\rm{d}}\tilde r.
\end{align}
Applying eq.~(\ref{eq19})-eq.~(\ref{eq20+}) and the dynamical apparent horizon of the FRW universe  $\tilde r$ to eq.~(\ref{eq19}), one gets the first Friedmann equation
\begin{align}
\label{eq21}
 - 4\pi G\left( {\rho  + p} \right) = \left( {\dot H - \frac{k}{{{a^2}}}} \right)f'(A),
\end{align}
where we set $\dot{\tilde r}= 0$ since the apparent horizon radius is fixed in an infinitesimal time interval. Next, substituting the continuity equation~(\ref{eq17}) into eq.~(\ref{eq21}), and then integrating, the result reads
\begin{align}
\label{eq22}
\frac{8}{3}\pi G\rho  =  - 4\pi \int {f'\left( A \right)\frac{{{\rm{d}}A}}{{{A^2}}}} ,
\end{align}
which is the second Friedmann equation. Now, by using eq.~(\ref{eq13}), the GUP corrected Friedmann equations evolves as
\begin{align}
\label{eq23}
- 4\pi G\left( {\rho  + p} \right) = \left( {\dot H - \frac{k}{{{a^2}}}} \right)\sqrt {1 + \frac{{\pi \beta }}{{2A}}} ,
\end{align}
and
\begin{align}
\label{eq24}
\frac{8}{3}\pi G\rho  = \frac{2}{{3\beta }}{\left( {4 + \frac{{2\pi \beta }}{A}} \right)^{3/2}} + \mathcal{C},
\end{align}
where $\mathcal{C}$  refers to an integration constant, which specific expression is determined by the boundary conditions in the vacuum energy dominated era \cite{cha29+}. When $A$  goes to infinity, the energy density becomes cosmological constant (i.e., $\rho  = \Lambda $), which leads to  $\mathcal{C} = {{8\pi G\Lambda } \mathord{\left/ {\vphantom {{8\pi G\Lambda } 3}} \right. \kern-\nulldelimiterspace} 3} - {{16} \mathord{\left/ {\vphantom {{16} {3\beta }}} \right. \kern-\nulldelimiterspace} {3\beta }}$. Then, inserting this expression of integration constant into eq.~(\ref{eq24}) and considering  $A = 4\pi {\tilde r^2} = {{4\pi } \mathord{\left/ {\vphantom {{4\pi } {\left( {{H^2} + {k \mathord{\left/ {\vphantom {k {{a^2}}}} \right. \kern-\nulldelimiterspace} {{a^2}}}} \right)}}} \right. \kern-\nulldelimiterspace} {\left( {{H^2} + {k \mathord{\left/{\vphantom {k {{a^2}}}} \right. \kern-\nulldelimiterspace} {{a^2}}}} \right)}}$, one yields
\begin{subequations}
\label{eq25}
\begin{align}
\label{eq25-1}
- 4\pi G\left( {\rho  + p} \right) & = \left( {\dot H - \frac{k}{{{a^2}}}} \right)\sqrt {1 + \frac{1}{8}\left( {{H^2} + \frac{k}{{{a^2}}}} \right)\beta } ,
 \\
\label{eq25-2}
\frac{8}{3}\pi G\left( {\rho  + \Lambda } \right) & = \frac{2}{{3\beta }}{\left[ {4 + \frac{1}{2}\left( {{H^2} + \frac{k}{{{a^2}}}} \right)\beta } \right]^{3/2}} - \frac{{16}}{{3\beta }}.
\end{align}
\end{subequations}
According to the research goal, we should study a flat universe that dominated by radiation. Hence, the tiny observed cosmological constant   $\Lambda$ and the spatial curvature constant $k$  can be ignored \cite{cha29+}. This argument implies that
\begin{subequations}
\label{eq26}
\begin{align}
\label{eq26-1}
 &- 4\pi G\left( {\rho  + p} \right)  = \dot H\sqrt {1 + \frac{{{H^2}}}{8}\beta } ,
\\
 &\frac{8}{3}\pi G\rho  = \frac{2}{{3\beta }}{\left( {4 + \frac{{{H^2}}}{2}\beta } \right)^{3/2}} - \frac{{16}}{{3\beta }}.
\end{align}
\end{subequations}
Clearly, the above two equations are the GUP corrected Friedmann equations which show the changes of energy density and pressure in the early Universe. When $\beta  \to 0$  the original Friedmann equations are recovered.
\subsection{GUP corrected baryon asymmetry factor}
\label{sec3.3}
For investigating the deviation of thermal equilibrium of the Universe, it is necessary to re-express the energy density and pressure of Friedmann equations as $\rho  = {\rho _0} + \delta \rho $ and $p = {p_0} + \delta p$, where $\rho_0$  and $p_0$  represent the pressure and density at thermal equilibrium, respectively. Substituting $\rho$  and  $p$ into GUP corrected Friedmann equations~(\ref{eq26}), one has
\begin{subequations}
\label{eq27}
\begin{align}
\label{eq27-1}
{\rho _{{\rm{GUP}}}} & =  - \frac{2}{{G\pi \beta }} + \left( {\frac{2}{{G\pi \beta }} + \frac{2}{3}{\rho _0}} \right)\sqrt {1 + \frac{1}{3}G\pi \beta {\rho _0}}
\nonumber \\
& = {\rho _0} + \frac{{G\pi {\rho ^2}\beta }}{{12}} - \frac{{{G^2}{\pi ^2}{\rho ^3}{\beta ^2}}}{{216}} + \mathcal{O}\left( {{\beta ^3}} \right),
\\
{p_{{\rm{GUP}}}} & = \frac{2}{{G\pi \beta }} + \left[ {\frac{{{\rho _0}}}{9}\left( {1 + 3w} \right) - \frac{2}{{3G\pi \beta }}} \right]\sqrt {9 + 3G\pi \beta {\rho _0}} .
\nonumber \\
& = w{\rho _0} + \frac{1}{{12}}G\pi \left( {1 + 2w} \right)\beta \rho _0^2 - \frac{1}{{216}}{G^2}{\pi ^2}\left( {2 + 3w} \right)\rho _0^3{\beta ^2} + \mathcal{O}\left( {{\beta ^3}} \right).
\end{align}
\end{subequations}
It is noteworthy that, in the early stages of the Universe, the effect of QG made the  ${\rho _{{\rm{GUP}}}} $  and ${p_{{\rm{GUP}}}}$  deviate from the cases at thermal equilibrium. As increases of the scale of the Universe, the effect of QG gradually weakens. So far, the QG effect in our universe is so weak that it can be ignored. Next, substituting eq.~(\ref{eq27}) into the trace of the Einstein equation~(\ref{eq5}), it follows that
\begin{align}
\label{eq28}
{R_{{\rm{GUP}}}}  = &  - 8\pi G\left( {{\rho _{{\rm{GUP}}}}   - 3{p_{{\rm{GUP}}}}} \right)
  \nonumber \\
 = &   \frac{8}{3}G\pi \left( {1 + 9w} \right){\rho _0}\sqrt {1 + \frac{1}{3}G\pi \beta {\rho _0}}  +  \frac{{64}}{{3\beta }}\left( {3 - \sqrt {9 + 3G\pi \beta {\rho _0}} } \right).
\end{align}
Applying the continuity equation~(\ref{eq17}) and the second Friedmann equation in thermal equilibrium ${H^2} = {{8\pi G{\rho _0}} \mathord{\left/ {\vphantom {{8\pi G{\rho _0}} 3}} \right. \kern-\nulldelimiterspace} 3}$  to eq.~(\ref{eq28}), the modified time derivative of the Ricci scalar takes the following form
\begin{align}
\label{eq29}
{\dot R_{{\rm{GUP}}}}   =  - \frac{{8\sqrt 2 G{\pi ^{3/2}}\left( {1 + w} \right){\rho _0}\left[ {G\pi \beta {\rho _0} - 6 + 9w\left( {2 + G\pi \beta {\rho _0}} \right)} \right]}}{{\sqrt {\pi \beta  + {3 \mathord{\left/ {\vphantom {3 {G{\rho _0}}}} \right. \kern-\nulldelimiterspace} {G{\rho _0}}}} }}.
\end{align}
Here we set $w = {1 \mathord{\left/ {\vphantom {1 3}} \right. \kern-\nulldelimiterspace} 3}$  since the universe is in an era dominated by radiation. With this setting, eq.~(\ref{eq29}) can be rewritten as follows
\begin{align}
\label{eq30}
{\dot R_{{\rm{GUP}}}} =   - \frac{{128\sqrt 2 {G^2}{\pi ^{5/2}}\beta \rho _0^2}}{{3\sqrt {\pi \beta  + {3 \mathord{\left/ {\vphantom {3 {G{\rho _0}}}} \right. \kern-\nulldelimiterspace} {G{\rho _0}}}} }},
\end{align}
where the original case ${\dot R}=0$ is  obtained for  $\beta\rightarrow0$. Finally, substituting eq.~(\ref{eq30}) into eq.~(\ref{eq3}), the expression for BAF is modified as follows
\begin{align}
\label{eq31}
{\eta _{{\rm{GUP}}}} = {\left. {\frac{{160{g_b}{G^{5/2}}\beta \rho _0^{5/2}}}{{{g_*}M_*^2T}}\sqrt {\frac{{2\pi }}{{3 + G\pi \beta {\rho _0}}}} } \right|_{{T_D}}}.
\end{align}
Obviously, the modified BAF is not only related to the parameters from the standard result~(\ref{eq3}), but also to the deformation parameter  $\beta$. According to the previous discussion, the effect of GUP is active in the radiation domination era, so our result can effectively avoid the occurrence of non-zero BAF. Besides, it is found that the original result~(\ref{eq3}) osatisfies only the first two Sakharov conditions \cite{cha7}. However, by using the modified corrected entropy~(\ref{eq14}) and Friedmann equations~(\ref{eq26}), we broke the thermodynamic equilibrium and met the third Sakharov condition. Therefore, our result satisfies all three Sakharov conditions, which indicates the GUP can generate a sufficient baryon asymmetry for at the radiation domination era.

\section{Constraints for the GUP parameter $\beta_0$}
\label{sec4}
Theoretically, the GUP parameters are always assumed as 1 so that the effect of GUP is valid when the energy is close to the Planck scale. However, when ignoring this assumption, the bound of the GUP parameter can be estimated by experiment data and observation results. The research on the bounds of GUP parameters has always been one of the important focuses, which could help to understand the effect of QG in the theoretical realm \cite{chb49,chb50,chb51,chb52}. Therefore, in this section, we will confront the theoretical result~(\ref{eq31}) with the observational results to obtain the bound of the parameter of the high-order GUP. In order to achieve this goal, one needs to replace the gravitational constant to Planck mass (i.e.  $G = {1 \mathord{\left/ {\vphantom {1 {m_p^2}}} \right.  \kern-\nulldelimiterspace} {m_p^2}}$ with ${m_p} \sim 1.22 \times {10^{19}}{\rm{GeV}}$) and express the density at thermal equilibrium as  ${\rho _0} = {{\pi {g_*}{T^4}} \mathord{\left/ {\vphantom {{\pi {g_*}{T^4}} {30}}} \right. \kern-\nulldelimiterspace} {30}}$ \cite{cha4}, then the GUP corrected BAF can be rewritten as
\begin{align}
\label{eq32}
{\eta _{{\rm{GUP}}}} = \frac{{8{g_b}g_*^{3/2}{\pi ^3}T_D^9\beta }}{{45m_p^5M_*^2\sqrt {45 + {{{g_*}{\pi ^2}T_D^4\beta } \mathord{\left/ {\vphantom {{{g_*}{\pi ^2}T_D^4\beta } {2m_p^2}}} \right. \kern-\nulldelimiterspace} {2m_p^2}}} }}.
\end{align}
By employing  $\beta  = {{{\beta _0}} \mathord{\left/ {\vphantom {{{\beta _0}} {m_p^2}}} \right. \kern-\nulldelimiterspace} {m_p^2}}$ and the values of  ${g_b} \sim \mathcal{O}\left( 1 \right)$,  ${g_*} \sim 106$, ${M_*} = {{{m_p}} \mathord{\left/ {\vphantom {{{m_p}} {\sqrt {8\pi } }}} \right. \kern-\nulldelimiterspace} {\sqrt {8\pi } }}$  and  ${T_D} \sim 2 \times {10^{16}}{\rm{GeV}}$ \cite{chb45}, then solving eq.~(\ref{eq32}), one has
\begin{align}
\label{eq33}
{\beta _0}  = 4.8 \times {10^{ - 82}}\left( {2.3 \times {{10}^{112}}\eta _{{\rm{GUP}}}^2} { + \sqrt {1.2 \times {{10}^{204}}\eta _{{\rm{GUP}}}^2 + 5.5 \times {{10}^{224}}\eta _{{\rm{GUP}}}^4} } \right).
\end{align}
Clearly, the above equation implies that the bound of $\beta_0$  is determined by the observed value of  ${\eta _{{\rm{GUP}}}}$. In the past forty years, many astronomical observations have given the range of  $\eta$, which can be used to constraint the region of the higher-order GUP. Now, according to the data in~\cite{cha3+,cha46,cha47,cha48,cha49}, the bounds on $\beta_0$  are shown in table~\ref{tab1}.

\begin{figure}[tbp]
\centering
\includegraphics[width=0.65
\textwidth]{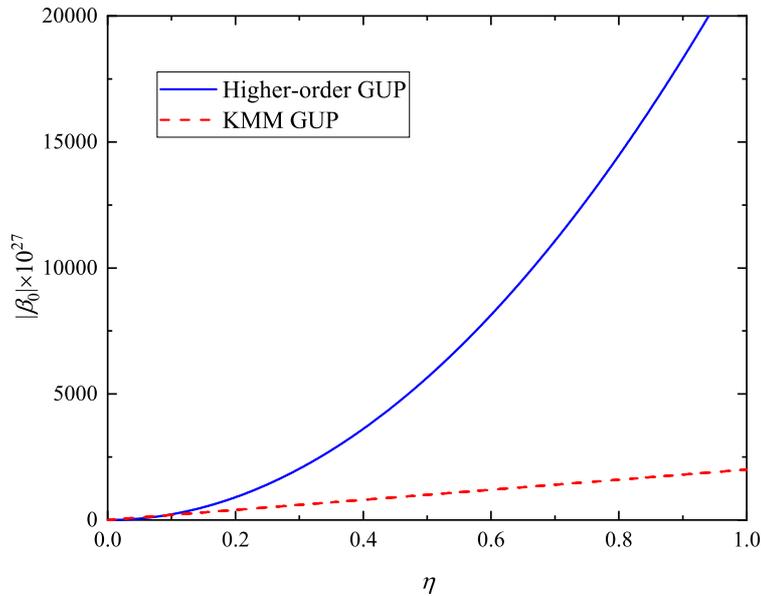}
\caption{\label{fig1} The effect of different GUP models on the  $\eta-{{\beta _0}}$ relationship.}
\end{figure}

\begin{table}[tbp]
\centering
\resizebox{\textwidth}{!}
{
\begin{tabular}{|c| c| c|}
\hline
Data source & $\eta$ & $\beta_0$\\
\hline
Planck observations   \cite{cha3+}                                                       &   $\eta  \le 6.2 \times {10^{ - 10}}$                                                           &    ${\beta _0} \le 8.7 \times {10^{12}}$   \\
Deuterium and Hydrogenium abundance \cite{cha46}                           &   $5.9 \times {10^{ - 10}} \le \eta  \le 6.3 \times {10^{ - 10}}$     &    $7.9 \times {10^{12}} \le {\beta _0} \le 9.0 \times {10^{12}}$   \\
Deuterium and 3He abundances \cite{cha47}        &  $5.7 \times {10^{ - 11}} \le \eta  \le 9.9 \times {10^{ - 11}}$     &   $8.4 \times {10^{10}} \le {\beta _0} \le 2.3 \times {10^{11}}$     \\
BBN  \cite{cha48}                                                               &    $3.4 \times {10^{ - 10}} \le \eta  \le 6.9 \times {10^{ - 10}}$    &    $2.6 \times {10^{12}} \le {\beta _0} \le 1.1 \times {10^{13}}$       \\
Acoustic peaks in CMB measured by WMAP \cite{cha49}                      &   $\eta  \le 6.3 \times {10^{ - 10}}$                      &    ${\beta _0} \le 9.0 \times {10^{12}}$      \\
Particle Data Group \cite{chx49}  & $\eta  \le 8.6 \times {10^{ - 11}}$ & ${\beta _0} \le 1.8 \times {10^{11}}$ \\
\hline
\end{tabular}
}
\caption{\label{tab1}  The data source, the ranges of $\eta$, and the bounds of  $\beta_0$.}
\end{table}
From table~\ref{tab1}, by employing data from different astronomical observations, one can see the bound  of $\beta_0$ is between $8.4 \times {10^{10}}  \sim 1.1 \times {10^{13}}$. Besides, the parameter of the higher-order GUP has a special relationship with that of the KMM model, that is,  ${\beta _0} \sim  - \beta _0^{{\rm{KMM}}}$. Therefore, comparing the results $\beta _0^{{\rm{KMM}}} = {{ - \eta } \mathord{\left/ {\vphantom {{ - \eta } {2.16 \times {{10}^{ - 19}}}}} \right. \kern-\nulldelimiterspace} {2.16 \times {{10}^{ - 19}}}}$  from~\cite{cha29+}, which shows $- 4.58 \times {10^8} \le \beta _0^{{\rm{KMM}}} \le  - 2.64 \times {10^8}$ (Deuterium and 3He abundances), it is easy to find that our results  $\beta_0$ are 2-5 orders of magnitude weaker than the absolute value of those of the KMM model  $\left| {\beta _0^{{\rm{KMM}}}} \right|$. In order to further study the influence of the two models on the ``$\eta-{{\beta _0}}$" relationship, we plot fig.~\ref{fig1}. One can see that the rad dashed line illustrates the KMM model case increases monotonically with  $\left| {\beta _0^{{\rm{KMM}}}} \right|$ with  a constant slope, whereas the higher-order GUP case presents a blue parabola with an increasing slope. As we stated previously, the higher-order GUP would reduces to the KMM model for  $\sqrt \beta  \Delta p \ll 1$. Therefore, we conclude that the above differences are caused by higher-order terms ${\cal O}\left( {{\beta _0}} \right)$ of GUP.

Furthermore, with the continuous improvement of astronomical observation equipment, one can now use gravitational wave signals to constrain the effect of QG at cosmological scales. For example, Nishizawa \cite{chx52} tested effect of QG with the gravitational-wave propagation and the modified photonic dispersion relation. Inspired by that work, we can constrain the parameter of the higher-order GUP models via the primordial gravitational wave experiments with space interferometers (such as LIGO, VIRGO, and KAGRA), those works would lay the foundation for understanding quantum gravity from a multi-messenger era perspective \cite{chx53}.

\section{Conclusion}
\label{sec5}
In the current work, the baryon asymmetry in the early Universe is studied from the perspective of thermodynamics and the effect of higher-order GUP. We started our investigation by analyzing the original gravitational baryogenesis. It is found that, although the original mechanism has coupled the spacetime and the baryon current, it still cannot explain the baryon asymmetry well since the Universe is in thermal equilibrium, which leads to a zero BAF and makes the third Sakharov condition unsatisfied. For fixing this problem, we derived the modified Bekenstein-Hawking entropy by employing a new higher-order GUP together with the holographic principle. Subsequently, according to this modification and the first law of thermodynamics,
we obtained the GUP corrected Friedmann equations. It is found that the modified Friedmann equations are in non-thermal equilibrium, which satisfies the third Sakharov condition and leads to $\eta  \ne 0$  in the radiation-dominated era. Those results showed that the effect of GUP can break thermal equilibrium and provide a feasible scheme for explaining baryon asymmetry in a radiation dominated Universe. Finally, using the different astronomical observations data, we constrained the bounds of the GUP parameter  $\beta _0$, whose bound between $8.4 \times {10^{10}}  \sim 1.1 \times {10^{13}}$. It is found that our results  the parameter of High-order GUP $\beta_0$ are 2-5 orders of magnitude weaker than the absolute value of that of KMM model $\left| {\beta _0^{{\rm{KMM}}}} \right|$. This indicates the bound of GUP parameter is depended on the its model, and the high-order terms ${\cal O}\left( {{\beta _0}} \right)$ of GUP  have an important impact on the baryon asymmetry.  In a word, our study provides the possibility to analyze the QG effect and its role in the early evolution of the Universe.

It should be noted that many new higher-order GUP \cite{cha33,cha34,cha35,chc36,cha36} and extended uncertainty principle models \cite{cha50,cha51} have been proposed recently. By combining them with baryon asymmetry could help us have a deeper understanding of the Universe. In addition, the CTA will start to reveal whether quantum gravity eventually plays a role in the Universe's dynamics, via photon dispersion relations \cite{cha52}, the data released by this observations will be important for our analysis of QG effects. We hope to address these issues in the future works.

\acknowledgments
The authors thank Guansheng He and the the anonymous referees for helpful suggestions and enlightening comments, which helped to improve the quality of this paper. This work was supported by the National Natural Science Foundation of China (Grant Nos. 12105231 and 11847048), the Guiding Local Science and Technology Development Projects by the Central Government of China (Grant No. 2021ZYD0031), the Sichuan Youth Science and Technology Innovation Research Team(Grant No. 21CXTD0038), and the Fundamental Research Funds of China West Normal University (Grant No. 20B009).



\begin{thebibliography}{99}
\bibitem{cha1}
C.D. Anderson, \emph{The Positive Electron}, \href{https://http://dx.doi.org/10.1103/PhysRev.43.491} {\emph{Phys. Rev.} \textbf{43} (1933) 491}.

\bibitem{chb1}
M.~Goldhaber, \emph
{Speculations on cosmogony}, \href{https://http://dx.doi.org/10.1126/science.124.3214.218} {\emph{Science} \textbf{124} (1956) 218}.

\bibitem{cha2}
C.L. Bennett, \emph{et al}. [WMAP Collaboration], \emph{First Year Wilkinson Microwave Anisotropy Probe (WMAP) Observations: Preliminary Maps and Basic Results},   \href{https://doi.org/10.1086/377253} {\emph{Astrophys. J. Suppl.} \textbf{148} (2003) 1} \href{https://arxiv.org/abs/astro-ph/0302207} {[arXiv:astro-ph/0302207
]}.

\bibitem{cha3}
S. Burles, K.M. Nollett and M. S. Turner, \emph{What Is The BBN Prediction for the Baryon Density and How Reliable Is It?} \href{https://doi.org/10.1103/PhysRevD.63.063512} {\emph{Phys. Rev. D} \textbf{63} (2001) 063512}  \href{https://arxiv.org/abs/astro-ph/0008495} {[arXiv:astro-ph/0008495]}.

\bibitem{cha3+}
 N. Aghanim, \emph{et al}. [Planck Collaboration],  \emph{Planck 2018 results. VI. Cosmological parameters},  \href{https://doi.org/10.1051/0004-6361/201833910} {\emph{Astron. Astrophys.} \textbf{641} (2020) A6} \href{https://arxiv.org/abs/1807.06209} {[arXiv:1807.06209]}.

\bibitem{cha4}
E.W. Kolb and M.S. Turner,  \emph{The early Universe}, \href{https://doi.org/10.1038/294521a0} {\emph{Nature}  \textbf{69} (1990) 1}.

\bibitem{cha5}
A.D. Sakharov, \emph{Violation of CP Invariance, C Asymmetry, and Baryon Asymmetry of the Universe}, \href{https://doi.org/10.1070/PU1991v034n05ABEH002497} {\emph{JETP Lett.} \textbf{5} (1967) 24.}


\bibitem{cha6}
L. Canetti, M. Drewes and M. Shaposhnikov, \emph{Matter and Antimatter in the Universe}, \href{
https://doi.org/10.1088/1367-2630/14/9/095012} {\emph{New J. Phys.} \textbf{14} (2012) 095012} \href{https://arxiv.org/abs/1204.4186} {[arXiv:1204.4186]}.

\bibitem{cha11}
G. Lambiase, S. Mohanty and A.R. Prasanna, \emph{Neutrino coupling to cosmological background: A review on gravitational Baryo/Leptogenesis},\href{https://doi.org/10.1142/S0218271813300309} {\emph{Int. J. Mod. Phys. D} \textbf{22} (2013) 1330030}. \href{https://arxiv.org/abs/1310.8459} {[arXiv:1310.8459]}.

\bibitem{chb11}
Y. Cui, \emph{A Review of WIMP Baryogenesis Mechanisms}, \href{
https://doi.org/10.1142/S0217732315300281} {\emph{Mod. Phys. Lett. A} \textbf{30} (2015) 1530028} \href{https://arxiv.org/abs/1510.04298} {[arXiv:1510.04298]}.

\bibitem{cha7}
H. Davoudiasl, R. Kitano, G.D. Kribis, H. Murayama and P. Steinhardt, \emph{Gravitational Baryogenesis}, \href{https://doi.org/10.1103/PhysRevLett.93.201301} {\emph{Phys. Rev. Lett.} \textbf{93} (2004) 201301} \href{https://arxiv.org/abs/hep-ph/0403019} {[arXiv:hep-ph/0403019]}.

\bibitem{cha8}
S.H.S. Alexander, M.E. Peskin and M.M. Sheikh-Jabbari, \emph{Leptogenesis from Gravity Waves in Models of Inflation}, \href{https://doi.org/10.1103/PhysRevLett.96.081301} {\emph{Phys. Rev. Lett.} \textbf{96} (2006) 081301} \href{https://arxiv.org/abs/hep-th/0403069} {[arXiv:hep-ph/0403019]}.

\bibitem{cha9}
H.M. Sadjadi, \emph{A Note on Gravitational Baryogenesis}, \href{https://doi.org/10.1103/PhysRevD.76.123507} {\emph{Phys. Rev. D} \textbf{84} (2011) 023509} \href{https://arxiv.org/abs/0709.0697} {[arXiv:0709.0697]}.

\bibitem{cha10}
G. Lambiase and S. Mohanty, \emph{Leptogenesis by curvature coupling of heavy neutrinos}, \href{https://doi.org/10.1103/PhysRevD.84.023509} {\emph{Phys. Rev. D} \textbf{84} (2011) 023509} \href{https://arxiv.org/abs/1107.1213} {[arXiv:1107.1213]}.

\bibitem{cha12}
G. Lambiase, S. Mohanty and A.R. Prasanna, \emph{Neutrino coupling to cosmological background: A review on gravitational Baryo/Leptogenesis}, \href{https://doi.org/10.1142/S0218271813300309} {\emph{Int. J. Mod. Phys. D} \textbf{22} (2013) 1330030} \href{https://arxiv.org/abs/1310.8459} {[arXiv:1310.8459]}.

\bibitem{cha13}
J.I. McDonald and G.M. Shore, \emph{Gravitational leptogenesis, C, CP and strong equivalence}, \href{https://doi.org/10.1007/JHEP02(2015)076} {\emph{JHEP} \textbf{1502} 076 (2015)} \href{https://arxiv.org/abs/1411.3669} {[arXiv:1411.3669]}.

\bibitem{chb14}
H. Li, M. Li and X. Zhang, \emph{Gravitational Leptogenesis and Neutrino Mass Limit}, \href{https://doi.org/10.1103/PhysRevD.70.047302} {\emph{Phys. Rev. D} \textbf{70} (2004) 047302} \href{https://arxiv.org/abs/hep-ph/0403281} {[arXiv:hep-ph/0403281]}.

\bibitem{cha14}
S.D. Odintsov and V.K. Oikonomou, \emph{Gauss-Bonnet Gravitational Baryogenesis}, \href{https://doi.org/10.1016/j.physletb.2016.06.074} {\emph{Phys. Lett. B} \textbf{760} (2016) 259} \href{https://arxiv.org/abs/1607.00545} {[arXiv:1607.00545]}.

\bibitem{chx13}
V.K. Oikonomou and  E.N. Saridakis, \emph{f(T)  gravitational baryogenesis}, \href{https://doi.org/10.1103/PhysRevD.94.124005}    {\emph{Phys. Rev. D} \textbf{94} (2016) 124005} \href{https://arxiv.org/abs/1607.08561} {[arXiv:1607.08561]}.

\bibitem{chx14}
V.K. Oikonomou, \emph{Constraints on singular evolution from gravitational baryogenesis}, \href{https://doi.org/10.1142/S021988781650033X} {\emph{Int. J Geom. Methods. M} \textbf{13} (2016) 1650033} \href{https://arxiv.org/abs/1512.04095} {[arXiv:1512.04095]}.

\bibitem{cha16}
K. Nozari and F. Rajabi, \emph{Baryogenesis in f(R,T) Gravity}, \href{https://dx.doi.org/10.1088/0253-6102/70/4/451} {\emph{Commun. Theor. Phys.} \textbf{70} (2018) 451}.

\bibitem{cha17}
E.H. Baffou, M.J.S. Houndjo, D.A. Kanfon and I.G. Salako, \emph{f(R,T) models applied to Baryogenesis}, \href{https://doi.org/10.1140/epjc/s10052-019-6559-0} {\emph{Eur. Phys. J.} C \textbf{79} (2019) 112} \href{https://arxiv.org/abs/1808.01917} {[arXiv:1808.01917]}.

\bibitem{cha18}
K. Atazadeh, \emph{Gravitational baryogenesis in DGP brane cosmology}, \href{https://doi.org/10.1140/epjc/s10052-018-5952-4} {\emph{Eur. Phys. J. C} \textbf{78}, (2018) 455} \href{https://arxiv.org/abs/2012.12965} {[arXiv:2012.12965]}.

\bibitem{chxx19}
V.~Antunes, I.~Bediagab and Mario Novelloa, \emph{Gravitational baryogenesis without CPT violation}, \href{https://doi.org/10.1088/1475-7516/2019/10/076} {\emph{JCAP}  \textbf{03} (2019) 013}  \href{https://arxiv.org/abs/1909.03034} {[arXiv:1909.03034]}.

\bibitem{chb18}
S. Bhattacharjee, \emph{Gravitational baryogenesis in extended teleparallel theories of gravity} \href{https://doi.org/10.1016/j.dark.2020.100612} {\emph{Phys. Dark Universe} \textbf{30} (2020) 100612} \href{https://arxiv.org/abs/2005.05534} {[arXiv:2005.05534]}.

\bibitem{cha19}
M. Fukushima, S. Mizuno and K. Maeda, \emph{Gravitational Baryogenesis after Anisotropic Inflation}, \href{https://doi.org/10.1103/PhysRevD.93.103513} {\emph{Eur. Phys. J. C} \textbf{93} (2016) 103513} \href{https://arxiv.org/abs/1603.02403} {[arXiv:1603.02403]}.

\bibitem{chx15}
S.D. Odintsov and V.K. Oikonomou, \emph{Loop quantum cosmology gravitational baryogenesis}, \href{https://doi.org/10.1209/0295-5075/116/49001} {\emph{EPL} \textbf{116} (2016) 49001} \href{https://arxiv.org/abs/1610.02533} {[arXiv:1610.02533]}.

\bibitem{chb19}
N. Smyth, L. Santos-Olmsted and S. Profumo,  \emph{Gravitational Baryogenesis and Dark Matter from Light Black Holes}, \href{https://doi.org/10.1088/1475-7516/2022/03/013} {\emph{JCAP} \textbf{03} (2022) 013} \href{https://arxiv.org/abs/2110.14660} {[arXiv:2110.14660]}.

\bibitem{cha20}
M. Maggiore, \emph{The algebraic structure of the generalized uncertainty principle}, \href{https://doi.org/10.1016/0370-2693(93)90785-G} {\emph{Phys. Lett. B} \textbf{319} (1993) 83} \href{https://arxiv.org/abs/hep-th/9309034} {[arXiv:hep-th/9309034]}.

\bibitem{cha21}
G. Amelino-Camelia, \emph{Relativity in space-times with short-distance structure governed by an observer-independent (Planckian) length scale}, \href{https://doi.org/10.1142/S0218271802001330} {\emph{Int. J. Mod. Phys. D} \textbf{11} (2002) 35} \href{https://arxiv.org/abs/gr-qc/0012051}  {[arXiv:gr-qc/0012051]}.

\bibitem{cha22}
F. Scardigli, \emph{Generalized Uncertainty Principle in Quantum Gravity from Micro-Black Hole Gedanken Experiment},  \href{https://doi.org/10.1016/S0370-2693(99)00167-7} {\emph{Phys. Lett. B} \textbf{452} (1999) 39} \href{https://arxiv.org/abs/hep-th/9904025} {[arXiv:hep-th/9904025]}.

\bibitem{cha23}
W. Chemissany, S. Das, A.F.  Ali and E.C. Vagenas, \emph{Effect of the Generalized Uncertainty Principle on Post-Inflation Preheating}, \href{https://doi.org/10.1088/1475-7516/2011/12/017}  {\emph{JCAP} \textbf{12} (2011) 017} \href{https://arxiv.org/abs/1111.7288}  {[arXiv:1111.7288]}.

\bibitem{cha24}
P. Chen, Y.C. Ong and D.-H. Yeom, \emph{Black Hole Remnants and the Information Loss Paradox},  \href{https://doi.org/10.1016/j.physrep.2015.10.007} {\emph{Phys. Rep.} \textbf{603} (2015) 1} \href{https://arxiv.org/abs/1412.8366} {[arXiv:1412.8366]}.

\bibitem{cha25}
H. Moradpour, A.H. Ziaie, S. Ghaffari and F. Feleppa, \emph{The generalized and extended uncertainty principles and their implications on the Jeans mass}, \href{https://doi.org/10.1093/mnrasl/slz098}  {\emph{Mon. Not.  R  Astron. Soc.} \textbf{488} (2019) L69} \href{https://arxiv.org/abs/1907.12940}  {[arXiv:1907.12940]}.

\bibitem{cha26}
Z.-W. Feng, H.-L. Li, X.-T. Zu and S.-Z. Yang, \emph{Quantum corrections to the thermodynamics of Schwarzschild-Tangherlini black hole and the generalized uncertainty principle}, \href{https://doi.org/10.1140/epjc/s10052-016-4057-1} {\emph{Eur. Phys. J. C} \textbf{76} (2016) 212} \href{https://arxiv.org/abs/1604.04702} {[arXiv:1604.04702]}.

\bibitem{cha27}
R. Casadio and F. Scardigli, \emph{Generalized Uncertainty Principle, Classical Mechanics, and General Relativity},  \href{http://dx.doi.org/10.1016/j.physletb.2020.135558} {\emph{Phys. Lett. B} \textbf{807} (2020) 135558}.

\bibitem{cha28}
A. Iorio, G. Lambiase, P. Pais and F. Scardigli, \emph{Generalized uncertainty principle in three-dimensional gravity and the BTZ black hole}, \href{http://dx.doi.org/10.1103/PhysRevD.101.105002} {\emph{Phys. Rev. D} \textbf{101} (2020) 105002}.

\bibitem{cha29}
D. Park and E. Jung, \emph{GUP and Point Interaction},  \href{https://doi.org/10.1103/PhysRevD.101.066007}  {\emph{Phys. Rev. D} \textbf{101} (2020) 066007} \href{https://arxiv.org/abs/2001.02850}  {[arXiv:2001.02850]}.

\bibitem{cha29+}
S. Das, M. Fridman, G. Lambiase and E. C. Vagenas, \emph{Baryon Asymmetry from the Generalized Uncertainty Principle} \href{https://doi.org/10.1016/j.physletb.2021.136841}  {\emph{Phys. Lett. B} \textbf{842} (2022) 136841} \href{https://arxiv.org/abs/2107.02077}  {[arXiv:2107.02077]}.

\bibitem{cha30}
A.F. Ali, S. Das and E.C. Vagenas, \emph{Discreteness of space from the generalized uncertainty principle}, \href{https://doi.org/10.1016/j.physletb.2009.06.061} {\emph{Phys. Lett. B} \textbf{678} (2009) 497}.

\bibitem{cha31}
P. Pedram, \emph{A Higher Order GUP with Minimal Length Uncertainty and Maximal Momentum}, \href{https://doi.org/10.1016/j.physletb.2012.07.005}  {\emph{Phys. Lett. B} \textbf{714} (2012) 317} \href{https://arxiv.org/abs/1110.2999}  {[arXiv:1110.2999]}.

\bibitem{cha32}
P. Pedram, \emph{A higher order GUP with minimal length uncertainty and maximal momentum II: Applications}, \href{https://doi.org/10.1016/j.physletb.2012.10.059} {\emph{Phys. Lett. B} \textbf{718} (2012) 638}.

\bibitem{cha33}
H. Shababi and W.S. Chung, \emph{On the two new types of the higher order GUP with minimal length uncertainty and maximal momentum
},   \href{https://doi.org/10.1016/j.physletb.2017.05.015} {\emph{Phys. Lett. B} \textbf{770} (2017) 445}.

\bibitem{cha34}
W.S. Chung and H. Hassanabadi, \emph{A new higher order GUP: one dimensional quantum system}, \href{https://doi.org/10.1140/epjc/s10052-019-6718-3} {\emph{Eur. Phys. J. C} \textbf{79} (2019) 213}.

\bibitem{cha35}
H. Hassanabadi, E. Maghsoodi and W.S. Chung, \emph{Analysis of black hole thermodynamics with a new higher order generalized uncertainty principle}, \href{https://doi.org/10.1140/epjc/s10052-019-6871-8} {\emph{Eur. Phys. J. C} \textbf{79} (2019) 358}.

\bibitem{chc36}
Z.-W. Feng, G. He, X. Zhou, X.-L. Mu and S.-Q. Zhou, \emph{Higher-order generalized uncertainty principle corrections to the Jeans mass}, \href{https://doi.org/10.1140/epjc/s10052-021-09549-z}  {\emph{Eur. Phys. J. C} \textbf{81} (2012) 754} \href{https://arxiv.org/abs/2006.01698}  {[arXiv:2006.01698]}.

\bibitem{cha36}
L. Petruzziello, \emph{Generalized uncertainty principle with maximal observable momentum and no minimal length indeterminacy}, \href{https://doi.org/10.1088/1361-6382/abfd8f}  {\emph{Class. Quant. Grav.} \textbf{38} (2021) 135005} \href{https://arxiv.org/abs/2010.05896}  {[arXiv:2010.05896]}.

\bibitem{chb36}
Y.C. Ong, \emph{Generalized Uncertainty Principle, Black Holes, and White Dwarfs: A Tale of Two Infinities}, \href{https://doi.org/10.1088/1475-7516/2018/09/015} {\emph{JCAP} \textbf{1809} (2018) 015} \href{https://arxiv.org/abs/1804.05176} {[arXiv:1804.05176]}.

\bibitem{chc37}
L. Buoninfante, G.G. Luciano and L. Petruzziello, \emph{Generalized Uncertainty Principle and Corpuscular Gravity},  \href{https://doi.org/10.1140/epjc/s10052-019-7164-y} {\emph{Eur. Phys. J. C} \textbf{79} (2019) 663} \href{https://arxiv.org/abs/1903.01382} {[arXiv:1903.01382]}.

\bibitem{chb37}
P. Jizba, H. Kleinert and F. Scardigli, \emph{Uncertainty Relation on World Crystal and its Applications to Micro Black Holes}, \href{https://doi.org/10.1103/PhysRevD.81.084030} {\emph{Phys. Rev. D} \textbf{81} (2010) 084030} \href{https://arxiv.org/abs/0912.2253} {[arXiv:0912.2253]}.

\bibitem{chy1}
K.N. Abazajian, K. Arnold and J. Austermann, \emph{et al.} \emph{Inflation physics from the cosmic microwave background and large scale structure}, \href{https://doi.org/10.1016/j.astropartphys.2014.05.013} {\emph{Astropart. Phys.} \textbf{63} (2015) 55} \href{https://arxiv.org/abs/1309.5381} {[arXiv:1309.5381]}.

\bibitem{chy2}
A. Kaya, \emph{The imprint of primordial gravitational waves on the CMB intensity profile}, \href{https://doi.org/10.1016/j.physletb.2021.136353} {\emph{Phys. Lett. B} \textbf{817} (2021) 136353} \href{https://arxiv.org/abs/2105.02236} {[arXiv:2105.02236]}.

\bibitem{chy3}
Y.F. Cai and Y. Wang, \emph{Testing quantum gravity effects with latest CMB observations},  \href{https://doi.org/10.1016/j.physletb.2014.06.019} {\emph{Phys. Lett. B} \textbf{48} (2018) 1191} \href{https://arxiv.org/abs/1404.6672} {[arXiv:1404.6672]}.

\bibitem{chy4}
A. Kempf, \emph{Quantum Gravity, Information Theory and the CMB},  \href{https://doi.org/10.1007/s10701-018-0163-2} {\emph{Phys. Lett. B} \textbf{735} (2014) 108} \href{https://arxiv.org/abs/1803.01483} {[arXiv:1803.01483]}.

\bibitem{cha52}
F. Iocco, \emph{et al}. [CTA consortium], \emph{Probing Dark Matter and Fundamental Physics with the Cherenkov Telescope Array}, \href{https://arxiv.org/abs/2106.03582} {arXiv: 2106.03582}.

\bibitem{cha37}
L. Pizza, \emph{Baryo-Leptogenesis induced by modified gravities in the primordial Universe}, \href{https://arxiv.org/abs/1506.08321}  {arXiv:1506.08321}.

\bibitem{cha38}
P. Bargue\~{n}o, E.C. Vagenas, \emph{Semiclassical corrections to black hole entropy and the generalized uncertainty principle}, \href{https://doi.org/10.1016/j.physletb.2015.01.016}  {\emph{Phys. Lett. B} \textbf{742} (2015) 15} \href{https://arxiv.org/abs/1501.03256}  {[arXiv:1501.03256]}.

\bibitem{chb34}
J.D. Bekenstein, \emph{Black holes and entropy}, \href{https://doi.org/10.1103/PhysRevD.7.2333} {\emph{Phys. Rev. D} \textbf{7} (1973) 2333}.

\bibitem{chb38}
A.J.M. Medved and E.C. Vagenas, \emph{When conceptual worlds collide: The GUP and the BH entropy},  \href{https://doi.org/10.1103/PhysRevD.70.124021} {Phys. Rev. D \textbf{70}, (2004) 124021} \href{https://arxiv.org/abs/hep-th/0411022} {[arXiv:hep-th/0411022]}.

\bibitem{chb39}
L. Xiang and X.Q. Wen, \emph{Black hole thermodynamics with generalized uncertainty principle}, \href{https://doi.org/10.1088/1126-6708/2009/10/046} {\emph{JHEP} \textbf{0910} (2009) 046} \href{https://arxiv.org/abs/0901.0603} {[arXiv:0901.0603]}.

\bibitem{cha35+}
A. Awad and A. F. Ali, \emph{Minimal Length, Friedmann Equations and Maximum Density},  \href{https://doi.org/10.1007/JHEP06(2014)093}  {\emph{JHEP} \textbf{06} (2014) 093}  \href{https://arxiv.org/abs/1404.7825}  {[arXiv:1404.7825]}.

\bibitem{cha39}
Z.-W. Feng, S.-Z. Yang, H.-L. Li, X.-T. Zu,    \emph{Thermodynamic phase transition of a black hole in rainbow gravity},  \href{https://doi.org/10.1155/2018/5968284}  {\emph{Phys. Lett. B} \textbf{772} (2017) 737} \href{https://arxiv.org/abs/1708.06627}  {[arXiv:1708.06627]}.

\bibitem{cha40}
F. Scardigli, \emph{Glimpses on the Micro Black Hole Planck Phase},  \href{https://doi.org/10.1103/PhysRevD.75.084003} {\emph{Symmetry} \textbf{12} (2020) 1519}.

\bibitem{cha41}
R.-G. Cai, L.-M. Cao and Y.-P. Hu, \emph{Corrected Entropy-Area Relation and Modified Friedmann Equations},  \href{https://doi.org/10.1088/1126-6708/2008/08/090} {\emph{JHEP} \textbf{0808} (2008) 090} \href{https://arxiv.org/abs/0807.1232} {[arXiv:0807.1232]}.

\bibitem{cha42}
A. Sheykhi, \emph{Entropic Corrections to Friedmann Equations}, \href{https://doi.org/10.1103/PhysRevD.81.104011}  {\emph{Phys. Rev. D} \textbf{81} (2010) 104011} \href{https://arxiv.org/abs/1004.0627}  {[arXiv:1004.0627]}.

\bibitem{cha43}
A. S. Sefedgar, \emph{From the entropic force to the Friedmann equation in rainbow gravity},  \href{https://doi.org/10.1209/0295-5075/117/69001} {\emph{EPL} \textbf{81} (2017) 69001}.

\bibitem{cha44}
Z.-W. Feng and S.-Z. Yang, \emph{Rainbow gravity corrections to the entropic force},  \href{https://doi.org/10.1155/2018/5968284}  {\emph{Adv. High Energy Phys.} \textbf{2018} (2018) 5968284} \href{https://arxiv.org/abs/1708.08324}  {[arXiv:1708.08324]}.

\bibitem{cha45}
M. Akbar and R.-G. Cai, \emph{Thermodynamic Behavior of Friedmann Equation at Apparent Horizon of FRW Universe}, \href{https://doi.org/10.1103/PhysRevD.75.084003}  {\emph{Phys. Rev. D} \textbf{75} (2007) 084003} \href{https://arxiv.org/abs/hep-th/0609128}  {[arXiv:hep-th/0609128]}.

\bibitem{chb45}
W.H. Kinney, E.W. Kolb, A. Melchiorri and A. Riotto, \emph{Inflation model constraints from the Wilkinson Microwave Anisotropy Probe three-year data}, \href{https://doi.org/10.1103/PhysRevD.74.023502}  {\emph{Phys. Rev. D} \textbf{74} (2006) 023502} \href{https://arxiv.org/abs/astro-ph/0605338}  {[arXiv:astro-ph/0605338]}.

\bibitem{cha46}
R.H. Cyburt, \emph{Primordial Nucleosynthesis for the New Cosmology: Determining Uncertainties and Examining Concordance}, \href{https://doi.org/10.1103/PhysRevD.70.023505}  {\emph{Phys. Rev. D} \textbf{70} (2004) 023505} \href{https://arxiv.org/abs/astro-ph/0401091}  {[arXiv:astro-ph/0401091]}.

\bibitem{cha47}
A. Riotto, \emph{Theories of Baryogenesis}, \href{https://arxiv.org/abs/hep-ph/9807454}  {CERN-TH/98-204, Trieste, Italy, 29 June-17 July 1998}.

\bibitem{cha48}
W. M. Yao \emph{et al}. [Particle Data Group], \emph{Review of Particle Physics}, \href{https://doi.org/10.1088/0954-3899/33/1/001} {\emph{J. Phys. G} \textbf{33} (2006) 1}.

\bibitem{cha49}
J. Dunkley, E. Komatsu, M.R. Nolta, D.N. Spergel, D. Larson, G. Hinshaw, L. Page, C.L. Bennett, B. Gold, N. Jarosik, J.L. Weiland, M. Halpern, R.S. Hill, A. Kogut, M. Limon, S.S. Meyer, G.S. Tucker, E. Wollack and E.L. Wright, \emph{Five-Year Wilkinson Microwave Anisotropy Probe (WMAP) Observations: Likelihoods and Parameters from the WMAP data}, \href{https://doi.org/10.1088/0067-0049/180/2/306}  {\emph{Astrophys. J Suppl. S}  \textbf{180} (2009) 306} \href{https://arxiv.org/abs/0803.0586}  {[arXiv:0803.0586]}.

\bibitem{chx49}
M. Tanabashi \emph{et al}. [Particle Data Group collaboration], \emph{Review of particle physics},
\href{https://doi.org/10.1103/PhysRevD.98.030001} {\emph{Phys. Rev. D} \textbf{98} (2018) 030001}.

\bibitem{chb49}
S. Das and E.C. Vagenas, \emph{Planck scale effects on some low energy quantum phenomena}, \href{https://doi.org/10.1103/PhysRevLett.101.221301} {\emph{Phys. Rev. Lett.} \textbf{101}  (2008) 221301}.

\bibitem{chb50}
S.~Sen, S.~Bhattacharyya and S.~Gangopadhyay, \emph{Probing the generalized uncertainty principle through quantum noises in optomechanical systems}, \href{https://doi.org/10.1088/1361-6382/ac55ab} {\emph{Class. Quant. Grav.} \textbf{39} (2022) 075020} \href{https://arxiv.org/abs/2112.13682} {[arXiv:2112.13682]}.

\bibitem{chb51}
F.~Feleppa, H.~Moradpour, C.~Corda and S.~Aghababaei, \emph{Constraining the generalized uncertainty principle with neutron interferometry}, \href{https://doi.org/10.1209/0295-5075/ac1240}{\emph{EPL} \textbf{135} (2021) 40003} \href{https://arxiv.org/abs/2111.00253}{[arXiv:2111.00253]}.

\bibitem{chb52}
G.G.~Luciano, \emph{Primordial big bang nucleosynthesis and generalized uncertainty principle}, \href{https://arxiv.org/abs/2111.06000} {\emph{Eur. Phys. J. C} \textbf{81} (2021) 1086}.

\bibitem{chx52}
A. Nishizawa, \emph{Generalized framework for testing gravity with gravitational-wave propagation. I. Formulation}, \href{https://doi.org/10.1103/PhysRevD.97.104037} {\emph{Phys. Rev. D} \textbf{97} (2018) 104037} \href{https://arxiv.org/abs/1710.04825} {[arXiv:1710.04825]}.

\bibitem{chx53}
A. Addazi, J. Alvarez-Muniz, R. Alves Batista, \emph{et al}. \emph{Quantum gravity phenomenology at the dawn of the multi-messenger era\textemdash{A} review}, \href{https://doi.org/10.1016/j.ppnp.2022.103948} {\emph{Prog. Part. Nucl. Phys.} \textbf{125} (2022) 103948} \href{https://arxiv.org/abs/2111.05659}. {[arXiv:2111.05659]}.

\bibitem{cha50}
J. Gin\'{e} and G.G. Luciano, \emph{Modified inertia from extended uncertainty principle(s) and its relation to MoND}, \href{https://doi.org/10.1140/epjc/s10052-020-08636-x} {\emph{Eur. Phys. J. C} \textbf{80} (2020) 1039}.

\bibitem{cha51}
E. Maghsoodi, H. Hassanabadi and W.S. Chung, \emph{Effect of the new extended uncertainty principle on black hole thermodynamics}, \href{https://doi.org/10.1209/0295-5075/129/59001} {\emph{EPL} \textbf{129} (2020) 59001}.


% Please avoid comments such as "For a review'', "For some examples",
% "and references therein" or move them in the text. In general,
% please leave only references in the bibliography and move all
% accessory text in footnotes.

% Also, please have only one work for each \bibitem.


\end{thebibliography}
\end{document}